\documentclass[a4paper,11pt]{article}
\usepackage{pos}

\title{George, I and the curvaton}

\author*[a]{Konstantinos Dimopoulos}

\affiliation[a]{Cosnortium for Fundamental Physics, Physics Department, Lancaster University\\
 Lancaster LA1 4YB, United Kingdom}

\emailAdd{k.dimopoulos1@lancaster.ac.uk}

\abstract{George Lazarides was a pivotal collaborator and friend to me. We worked together on several projects, developing and exploiting the curvaton hypothesis, which was new at the time. This is a brief overview of our joint research.}

\FullConference{Proceedings of the Corfu Summer Institute 2025 "School and Workshops on Elementary Particle Physics and Gravity" (CORFU2025)\\
27 April - 28 September, 2025\\
Corfu, Greece\\}

\begin{document}
\maketitle

\section{Getting to know George}

Even though I did my first degree in Physics in the Aristotle University of Thessaloniki (AUTH), I had not met George but near the end of my Ph.D. studies in Cambridge, at the first COSMO conference in Ambleside in 1997, organised by Leszek Roszkowski, then of Lancaster University. This was because George had a position in the Faculty of Engineering in AUTH, while I did my undergraduate studies in the Physics Department of the Faculty of Sciences. After obtaining my Ph.D. I had to complete my compulsory military service, until September 1999. While I was in the military, the research landscape in cosmology had radically shifted. After the observations of Boomerang and Maxima, the cosmic string paradigm for structure formation has collapsed leaving inflation as the only alternative proposal. Much of my doctorate studies were in cosmic strings, which were big in the UK at the time. Simultaneously, entirely new directions had opened, namely braneworlds, which were utterly unknown to me. In this climate, I was frantically looking for a postdoc. George saved me from this predicament, by offering to host me as a fellow of the Greek Stake Scholarship Foundation (IKY). In December of 1999, I did find a postdoc employment in Valencia. I interrupted the IKY fellowship and left Thessaloniki and George.

In 2001 I got another postdoc offering by Oxford, to be based, as a continuous visitor, in Lancaster. There I worked with David Lyth, who had just put forward the curvaton idea \cite{Lyth:2001nq,Lyth:2002my}. I had not lost touch with George, so this was the time to start collaborating with him. I discussed the curvaton idea with George. He found it very interesting, so this is how our collaboration began.

\section{The curvaton mechanism}

The curvaton idea is that the field responsible for the large scale structure formation in the Universe is other than inflaton field. Indeed, all light scalar fields undergo particle production during inflation. The superhorizon spectrum of perturbations $\delta\sigma$ of one such spectator field can be the dominant contribution to the curvature perturbation, which is ultimately responsible for structure formation. In this case, the scalar field in question is called curvaton \cite{Lyth:2001nq,Lyth:2002my}.

\begin{figure*}[t]
		\vspace{-3cm}
		\centering
		\includegraphics[scale=0.8]{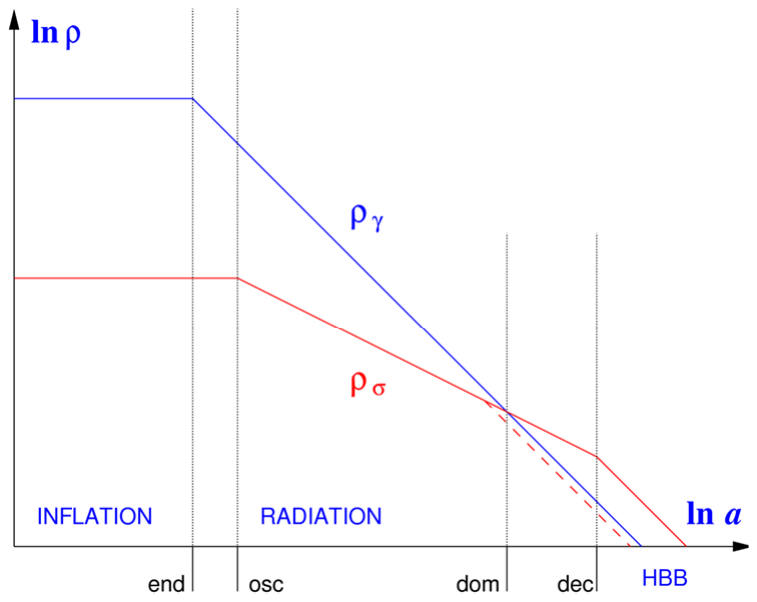}
\vspace{-12cm}
		\caption{Log-log plot of the evolution of the densities. The blue line corresponds to the density of the inflaton which becomes radiation at the end of inflation (prompt reheating is assumed). The red line corresponds to the density of the curvaton.}
		\label{curvaton}
	\end{figure*}

The curvaton mechanism operates as follows. During inflation, the curvaton field is frozen (overdamped) and remains subdominant. However, after the end of inflation, because $H(t)$ is decreasing, at some point
the curvaton mass becomes \mbox{$m>H(t)$}, which means that the curvaton unfreezes and begins oscillating around its vacuum expectation value. Such coherent oscillations correspond to massive particles (curvatons), whose density decreases with the Universe expansion as \mbox{$\rho_\sigma\propto a^{-3}$}. Thus, it can dominate the radiation background, which was the product on the inflaton decay and scales as \mbox{$\rho_r\propto a^{-4}$}. When it does so, it imposes its own curvature perturbation $\zeta_\sigma$ onto the Universe \mbox{$\delta\zeta=\hat\Omega_\sigma\zeta_\sigma$}, where \mbox{$\zeta_\sigma\sim\delta\sigma/\sigma$} and
\mbox{$\hat\Omega_\sigma\equiv
	\frac{3\Omega_\sigma}{4-\Omega_\sigma}$}, with \mbox{$\Omega_\sigma=\rho_\sigma/\rho$} being the curvaton density parameter \cite{Lyth:2002my}. The evolution in the curvaton scenario is depicted in Fig.~\ref{curvaton}.

As a result, the curvaton hypothesis liberates inflation model-building by removing responsibility for the curvature perturbation from the inflaton field
\cite{Dimopoulos:2002kt}. As such, it renders many theoretically well motivated inflation models naturally viable.

\section{Curvaton dynamics}

This is the first project we set up with George, along with David Lyth and a then PDRA with George, Roberto Ruiz de Austri \cite{Dimopoulos:2003ss}. We considered a curvaton field whose potential has both a mass-of-order $H$ term and a higher order non-renormalisable term. 
\begin{equation}
	V(\sigma,H)=\frac12(m^2\pm cH^2)\sigma^2+\lambda\frac{\sigma^{n+4}}{M^n}\,,
\end{equation}
where \mbox{$c={\cal O}(1)$} and $M$ is a large density scale. The contribution of order $H$ to the mass is naturally expected in supersymmetric theories \cite{Dine:1995uk}.

\begin{figure*}[t]
	\vspace{-2cm}
	\centering
	\includegraphics[scale=0.7]{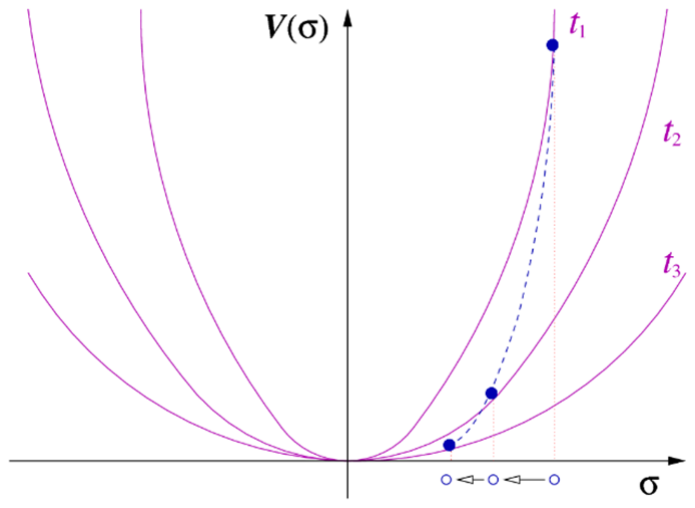}
	\vspace{-12cm}
	\caption{Schematic plot of the evolution of the potential \mbox{$V(\sigma)=\frac12 c H^2(t)\,\sigma^2$}. While the curvaton (depicted by a solid blue ball) rolls down the potential slope the potential opens up because $H(t)$ is decreasing.}
	\label{quasiquad}
\end{figure*}

We first considered that that the quasi-quadratic term \mbox{$V(\sigma)\simeq\frac12 cH^2\sigma^2$} is dominant. Then the density parameter of the curvaton is 
\begin{equation}
	\Omega_\sigma\propto a^{-\frac32(1-w)+2\sqrt{\mbox{max}\{0,c_x-c\}}}\,,
\end{equation}
where $w$ is the barotropic parameter of the background density and \mbox{$\sqrt{c_x}=\frac34(1-w)$}, with the curvaton oscillating when \mbox{$c>c_x$} or not when \mbox{$c<c_x$}, in which case it is just sliding down an opening-up potential, as shown in Fig.~\ref{quasiquad}.

Then we considered that the non-renormalisable term 
\mbox{$V(\sigma)\sim\lambda\sigma^{n+4}/M^n$} is dominant, in which case the curvaton density parameter is
\begin{equation}
	\Omega_\sigma\propto a^{-6(\frac{n+4}{n+6})+3(1+w)}\,,
\end{equation}
where \mbox{$n\leq n_c\equiv 2(\frac{1+3w}{1-w})$}. When \mbox{$n>n_c$},  we found an attractor solution that sends the curvaton perturbation $\delta\sigma$ to zero.

This work has amassed more than 120 inSPIRE citations to date and it was followed by another work of George with Roberto Trotta considering a mixed inflaton/curvaton scenario \cite{Lazarides:2004we}.

\section{The Peccei-Quinn field as curvaton}

The same set of collaborators produced also my second work with George \cite{Dimopoulos:2003ii}. This was based on the fact that the curvaton is not an ad hoc field; there exist many candidates in realistic beyond the standard model physics. In this spirit, we employed the Peccei-Quinn field, which solves the strong CP and the $\mu$ problems.

We considered the superpotential
\begin{equation}
	W=\frac{\lambda}{2m_P}N\bar N+\frac{\beta}{m_P}N^2h_1h_2\;,
\end{equation}
with the charges
\begin{center}
\begin{tabular}{lcccc}
PQ: & $N(-1)$ & $\bar N(1)$ & $h_1(1)$ & $h_2(1)$\\
R: & $N(1)$ & $\bar N(0)$ & $h_1(0)$ & $h_2(0)$
\end{tabular}.
\end{center}
For the $\mu$ term we have \mbox{$\mu\sim f_a^2/m_P\sim m_{3/2}\sim 1\,$TeV}, where 
\mbox{$f_a\sim\sqrt{m_{3/2}m_P}\sim 10^{11}\,$GeV} is the axion decay constant and $m_{3/2}$ is the gravitino mass.

With suitable rotations in field space we have \mbox{$\sigma=|N|,|\bar N|$}. Thus, we obtain the curvaton scalar potential
\begin{equation}
	V_{\rm PQ}(\sigma)=2\sigma^2m_{3/2}^2\left(1-2{\cal A}\lambda\frac{\sigma^2}{2m_{3/2}m_P}+\lambda^2\frac{\sigma^2}{m_{3/2}^2m_P^2}\right)\;,
\end{equation}
where \mbox{${\cal A}={\cal O}(1)$}. The curvaton vacuum expectation value (VEV) is
\begin{equation}
	\sigma_0=\frac{{\cal A}+\sqrt{{\cal A}^2-12}}{6\lambda}\,\sqrt{m_{3/2}m_P}=f_a\;.	
\end{equation}

\begin{figure*}[t]
	\vspace{-4cm}
	\centering
	\mbox{\hspace{-3cm}
	\includegraphics[scale=1]{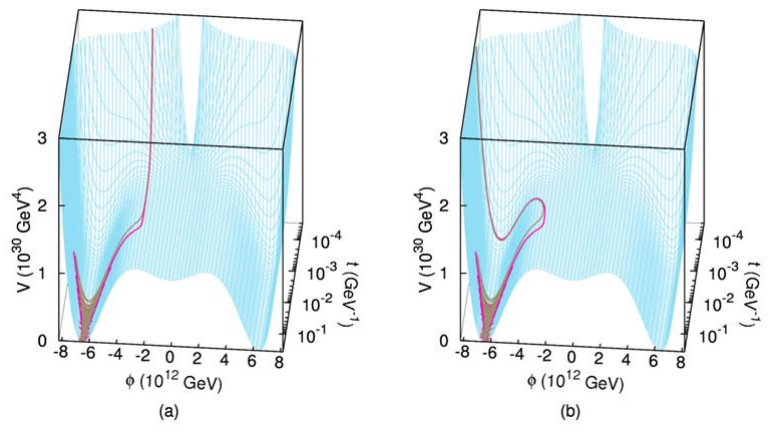}}
	\vspace{-18cm}
	\caption{Two possible trajectories in field space for the curvaton to experience tachyonic amplification of its perturbations when it loiters on top of a potential hill.}
	\label{loiter}
\end{figure*}

We found that the model works only if there is tachyon amplification of the PQ fluctuations. For this the 
PQ field must loiter on top of a local max of the scalar potential, as shown in Fig.~\ref{loiter}.

This work has amassed more than 90 inSPIRE citations to date.

\section{The orthogonal axion as curvaton}

\begin{figure*}[t]
	\vspace{-1cm}
	\centering
	\includegraphics[scale=0.6]{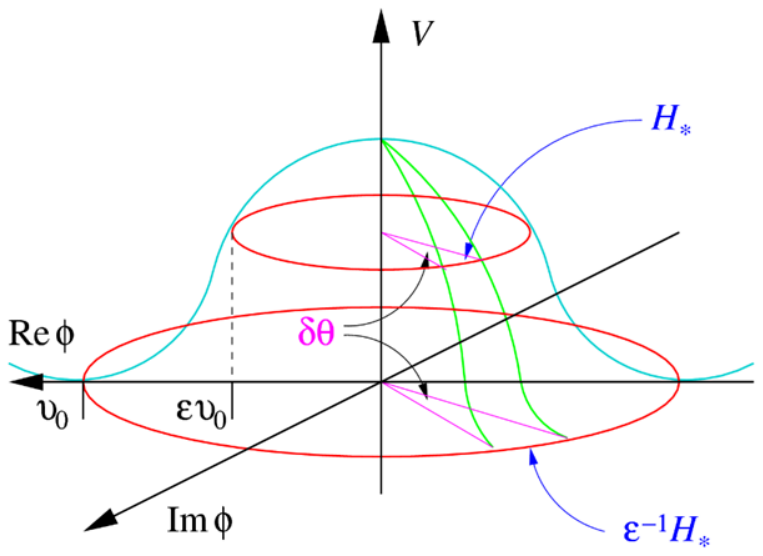}
	\vspace{-10cm}
	\caption{Schematic plot showing the enlargement of curvaton fluctuations, when the decay constant is enlarged from $\varepsilon v_0$ to $v_0$, with \mbox{$\varepsilon<1$}}
	\label{varax}
\end{figure*}

In Ref.~\cite{Lyth:2003dt}, it was shown that the simple curvaton model cannot accommodate low-scale inflation, \mbox{$H_*>10^7\,$GeV}. There may be one way out of this bound, namely considering as curvaton a PNGB with subsequent growth of its decay constant $f_a$. This enlarges the curvaton perturbations from their original value of $H_*/2\pi$ as shown in Fig.~\ref{varax}.

I explored this idea with the toy model \cite{Dimopoulos:2005bx}
\begin{equation}
	W=\frac{\lambda}{n+3}\frac{\phi^{n+3}}{m_P^n}\,.
\end{equation}
The minimum of the radial field $|\phi|$ during inflation is \mbox{$v_*\ll v_0$}, where \mbox{$v_0=f_a$} is the VEV. The scalar potential is \mbox{$V=(V|\phi|)+V(\sigma)$}, where
\begin{equation}
	V(|\phi|)=(c_\phi H^2-m_\phi^2)|\phi|^2+\lambda^2\frac{|\phi|^{2(n+2)}}{m_P^{2n}}\,,
\end{equation}
and
\begin{equation}
	V(\sigma)=(c_A H+A)\lambda\frac{v^{n+3}}{m_P^n}\left[1-\cos\left(\frac{\sigma}{v}\right)\right]\,,
\end{equation}
where \mbox{$c_\phi,c_A={\cal O}(1)$}. I also considered modular inflation with
\begin{equation}
	V(s)=V_{\rm inf}-\frac12 m_s^2s^2+\cdots\,,
\end{equation}
where \mbox{$m_s\sim m_{3/2}$}, with \mbox{$H_{\rm inf}\sim 1\,$TeV}\mbox{$\,\ll 10^7\,$TeV}. The model works for 
\begin{equation}
	m_\sigma(v_*)\ll m_\sigma(v_0)\sim\sqrt{Am_\phi}\sim m_{3/2}\sim 1 {\rm TeV}\,.
\end{equation}

With George, in our third project together, we constructed a much more realistic model, with superpotential~\cite{Dimopoulos:2005bp}
\begin{equation}
	W=\lambda\frac{P^{n+1} h_1 h_2}{m_P^n}+\sum_{k=0}^{\frac14(n+3)}\lambda_k\frac{S^{n+3-4k}(PQ)^{2k}}{m_P^n}\,,
\end{equation}
with the charges
\begin{center}
	\begin{tabular}{lcccc}
		PQ: & $P(-2)$ & $Q(2)$ & $S(0)$ & $h_1,h_2(n+1)$\\
		R: & $P(\frac{n+3}{2})$ & $Q(\frac{n-1}{2})$ & $S(\frac{n+1}{2})$ & $h_1,h_2(0)$
	\end{tabular}.
\end{center}
The model contains two PNGBs. One is the ultralight QCD axion. The other is the orthogonal axion, with \mbox{$m_\sigma\sim \hat A(v/v_0)$}, where \mbox{$\hat A=A+c_A H$} and \mbox{$A\sim m_{3/2}$}.

\begin{figure*}[t]
	\vspace{-2cm}
	\centering
	\includegraphics[scale=0.7]{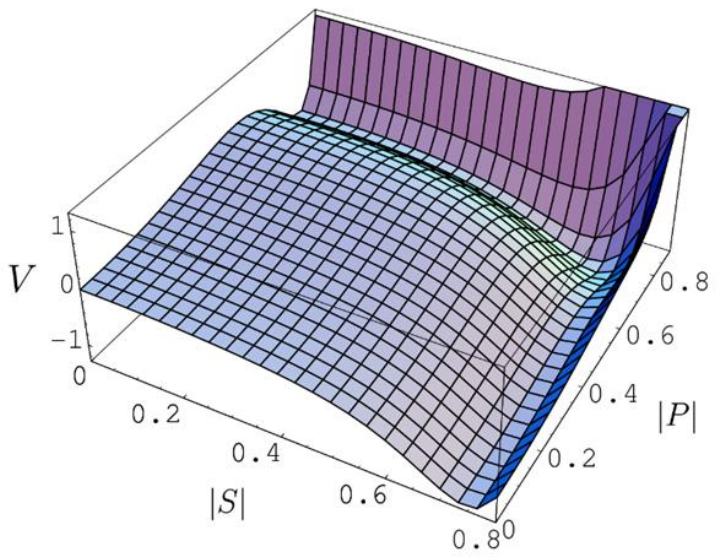}
	\vspace{-11cm}
	\caption{The F-term scalar potential features two valleys orthogonal between them.}
	\label{ortho}
\end{figure*}

The scalar potential features two valleys: The trivial valley, when \mbox{$S=0$}, and the shifted valley, when \mbox{$|S|\ll |P|\sim|Q|$} (see Fig.~\ref{ortho}). For the curvaton, the $P$ and $Q$ fields are at their VEVs and the potential becomes
\begin{equation}
	V(\sigma)\simeq \hat A^2\frac{v^4}{v_0^2}\left[1-\cos\left(\frac{\sigma}{v}\right)\right]\,.
\end{equation}
The model works for \mbox{$n=5,9$}. For \mbox{$n=5$}, the superpotential is 
\mbox{$W=m_P^{-5}[\lambda_0 S^8+\lambda_1 S^4(PQ)^2+\lambda_2(PQ)^4]$}.
Then, \mbox{$f_a=v_0\sim 10^{15.5}\,$GeV}.

\section{Vector curvaton}

In order to discuss the last project I worked with George I have to briefly mention the vector curvaton paradigm. I have introduced this in 2006 \cite{Dimopoulos:2006ms}. It was the first study of the contribution of vector fields to the curvature perturbation in the Universe. I employed a single Abelian vector boson field in a Proca theory of non-zero mass $m$. The energy-momentum tensor of the theory is
\begin{equation}
	T_\mu^\nu=\mbox{diag}(\rho_A, -p_\perp,-p_\perp,+p_\perp)\,,
\end{equation}
where the energy density and perpendicular pressure are
\begin{equation}
	\rho_A=\rho_{\rm kin}+V_A\quad\mbox{\rm and}\quad p_\perp=\rho_{\rm kin}-V_A\;,
\end{equation}
with
\begin{equation}
	\rho_{\rm kin}\equiv-\frac14 F_{\mu\nu}F^{\mu\nu}
	\quad\mbox{\rm and}\quad V_A\equiv-\frac12 m^2 A_\mu A^\mu\,.
\end{equation}

Originally, the vector field is frozen and subdominant during inflation. Eventually, after the end of inflation, as the Hubble parameter $H(t)$ decreases, we have \mbox{$m\sim H$} and the vector field thaws and begins undergoing harmonic oscillations, exactly as in the scalar curvaton scenario. During the oscillations \mbox{$\overline{\rho_{\rm kin}}\approx\overline{V_A}$},
which means that \mbox{$\overline{p_\perp}\approx 0$}. The oscillating vector curvaton behaves like isotropic, pressureless matter, whose density scales as \mbox{$\rho_A\propto a^{-3}$}. Thus, it eventually dominates the radiation background (with density \mbox{$\rho_r\propto a^{-4}$}), which is a product of the decay of the inflaton field. When this happens, the vector curvaton imposes onto the Universe its own curvature perturbation without introducing any anisotropy. Note that this curvature perturbation is scalar in nature. 

This proposal was put forward before the discovery of the Higgs field in CERN. Thus, no fundamental scalar field had been observed yet, so it was important to see whether one could generate the scalar curvature perturbation in the Universe without scalar fields.

The most promising vector curvaton model considered a theory where the gauge kinetic function is modulated by the inflaton field 
\mbox{$\delta{\cal L}=-\frac14f(\phi)F_{\mu\nu}F^{\mu\nu}$} \cite{Dimopoulos:2009am}. Then, it can be shown that, to obtain a scale invariant spectrum of vector field perturbations, one needs \mbox{$f\propto a^{-1\pm 3}$}.
Because the kinetic function is the inverse of the gauge coupling \mbox{$f\sim 1/g^2$} and at the end of inflation \mbox{$f=1$}, the only way to avoid a strongly coupled theory would be to consider \mbox{$f\propto a^{-4}$}. In Ref.~\cite{Wagstaff:2010qhd}, we have shown that this scaling is an attractor. 

The Klein-Gordon equation for the inflaton becomes
\begin{equation}
	\ddot\phi+3H\dot\phi+V'=-\frac14 f'F^2\equiv{\cal B}\,,
\end{equation} 
where the prime denotes derivative with respect to the inflaton field, and \mbox{$F^2=F_{\mu\nu}F^{\mu\nu}$}.
Therefore, the inflaton field experiences a "flattened" potential \mbox{$|V_{\rm eff}'|=|V'-{\cal B}|<|V'|$}.
This means that we could have steep inflation, where 
the variation of the inflaton is impeded primarily by the backreaction $\cal B$.

This offers a novel solution to the infamous $\eta$-problem of modeling inflation in supergravity.
$\eta$ is one of the slow-roll parameters, which typically is of order unity in supergravity, but slow-roll inflation requires \mbox{$|\eta|\ll 1$}, for inflation to last long enough, for the spectral index $n_s$ to be near unity as suggested by observations (\mbox{$n_s=1+2\eta+{\cal O}(\epsilon)$}) and for generating a superhorizon spectrum of inflaton field perturbations. All these issues are overcome with vector backreaction.

\section{\boldmath Overcoming the $\eta$-problem in supergravity hybrid inflation}

This is the fourth and final project with George \cite{Dimopoulos:2011ym}. My then doctorate student 
Jacques Wagstaff, did a six month secondment in Thessaloniki, with George. This is when we did this work. We considered the most general superpotential of the form
\begin{equation}
	W=\kappa S(\Phi\bar\Phi-M^2)+\mbox{non-renormalisable terms}\,,
\end{equation}
with $M$ being a large mass-scale,
and the K\"{a}hler potential
\begin{equation}
	K=|S|^2+\frac{\alpha}{4}\frac{|S|^4}{m_P^2}+\cdots\,,\label{K}
\end{equation}
where the ellipsis denotes even higher orders. Ignoring these and setting \mbox{$\alpha=0$} leaves the minimal K\"{a}hler potential \mbox{$K=|S|^2$}. In this case, an accidental cancellation eliminates $\eta$. However, in the general case shown in Eq.~\eqref{K}, \mbox{$\eta\simeq\alpha={\cal O}(1)$} and the model suffers from the $\eta$-problem.

To overcome the $\eta$-problem we considered the backreaction from a vector curvaton field, with gauge kinetic function given by
\begin{equation}
	f(S)=\exp\left[q\left(\frac{S}{M}\right)^n\right]
\end{equation} 
The model works when \mbox{$|q|\gtrsim 3$} and \mbox{$n\gtrsim 4$}. Then the scalar spectral index is 
\begin{equation}
	n_s=1-\frac{2(n-1)}{nN_*}\,,
\end{equation}
where \mbox{$N_*\simeq 60$} is the number of efolds of remaining inflation when the cosmological scales exit the horizon. For the range, \mbox{$3<n<+\infty$} we obtain \mbox{$0.978>n_s>0.967$} respectively. This is by far the most natural solution to the $\eta$-problem in supergravity. 

\section{Farewell}

Curvaton cosmology has been the meeting place for me and George. I knew George for more than 20 years  (we collaborated for more than a decade). He was a mentor and a friend to me. Whenever I returned to Thessaloniki, my home town, I always would pay him a visit in his office in the Faculty of Engineering. We would talk of physics but also about other things, about history and religion (George was deeply knowledgeable on these subjects), about little-known tsipouradika and mezedopoleia in Thessaloniki (we exchanged notes), about his rebetika evenings in Thasos. I have been in his house and knew his lovely wife Veta. When I got married he offered to give a tour to my English guests (Anne Davis, David Lyth and his wife Margaret) of the byzantine churches in Thessaloniki. He invited my wife and me to the wedding of his son. I confided in him, for example when I had a serious health problem. I cherished his advice. Cosmology and I will miss him…

\begin{figure*}[t]
	\vspace{-2cm}
	\centering
	\includegraphics[scale=0.7]{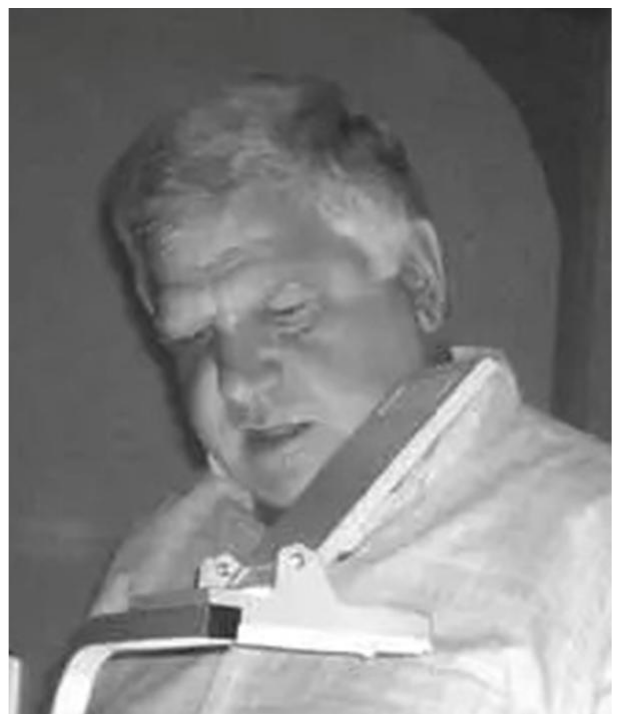}
	\vspace{-9cm}
	\caption{Professor George Lazarides delivering a presentation.}
	\label{LazaPhoto}
\end{figure*}

\end{document}